\documentstyle[preprint,revtex]{aps}
\begin{document}
\draft
\begin{title}
 Subgap Conductivity of a Superconducting-Normal Tunnel Interface
\end{title}
\author{F. W. J. Hekking and Yu.V. Nazarov}
\begin{instit}
 Institut f\"ur Theoretische Festk\"orperphysik, Universit\"at Karlsruhe,
 Postfach 6980, 7500 Karlsruhe, FRG\\
\end{instit}

\begin{abstract}
At low temperatures, the transport through a superconducting-normal
tunnel interface is due to tunneling of electrons in pairs.
The probability for this process is shown to depend on the layout
of the electrodes near the tunnel junction, rather than on
properties of the tunnel barrier.
This dependence is due to interference of the electron
waves on a space scale determined by the coherence length,
either in the normal or the superconducting metal.
The approach developed allows us to evaluate
the subgap current for different layouts of interest.
\end{abstract}

\pacs{PACS numbers: 74.20.Fg, 74.50 +r, 72.10 Fk}
It is well-known that the charge transport  through a tunnel NS interface
between a normal metal and a superconductor
 is strongly
suppressed at voltages lower than $\Delta/e$, $\Delta$ being the
superconducting
energy gap~\cite{Tinkham}. Indeed, energy conservation forbids the  transfer of
a normal electron
with an energy below the gap
to the superconductor, since it would have been converted into a quasiparticle
with an energy larger than
$\Delta$.

Experimentally, some residual conductivity has been observed at subgap voltages
even at very
low temperatures. There is a tendency to ascribe this either to imperfections
in the tunnel
barrier or to normal inclusions in the superconductor. Another mechanism of the
subgap
conductivity is the so-called {\it two-electron tunneling}~\cite{Wilkins}.
The point is that two normal electrons
can be converted into a Cooper pair, thus this transfer may cost no energy.
The current will be proportional to the fourth
power of tunnel matrix elements; therefore it is much smaller than the
one-electron current.

The problem was previously treated  under the simple assumption that the
electron wavefunctions
in both metals are just plane waves. In this case one can consider a barrier of
arbitrary
transparency in order to describe the crossover from a
tunnel to a perfectly conducting interface~\cite{Blonder}.
But some important physics may be missed under this assumption. Let
us compare the two realizations of the NS interface depicted in Fig. 1.
In Fig. 1a
the electron transmitted through the interface does not experience any
scattering in the metals
and never gets back to the junction. The plane wave picture seems to be
relevant for such
a geometry. An alternative situation is shown in Fig. 1b. This case is usually
realized
when the tunnel junction is formed by imposing two thin metal films. The
transmitted electron
gets back to the junction many times before leaving the junction region. Thus
the tunneling
occurs between electron states of very complex structure which emerges from
interference between
scattered waves.

This interference has no effect on one-electron transport, since the average
one-electron density of states
does not depend on the stucture of the wave function. However, it matters for
two-electron
tunneling, since two electrons penetrating the barrier will  interfere.
Such an interference occurs at a space scale corresponding to the energy
difference between
the two electron states. It makes the probability of two-electron tunneling
dependent on the system
layout at the corresponding mesoscopic space scale.
Our aim is to evaluate this interference effect for an arbitrary given layout.
The rapid progress of nanotechnology makes it possible to fabricate numerous
relevant
stuctures, so it is worthwile to be able to give guidelines to a designer.
As we will see below, the subgap conductivity is strongly enhanced if the
interference effect is
essential.

We first review  shortly the two-electron tunneling through
a superconducting-normal interface
as it has been
discussed by Wilkins~\cite{Wilkins} and more recently by Hekking {\sl et
al.}~\cite{Hekking}.
The total Hamiltonian can be written as
$
	\hat{H}
	=
	\hat{H}_N + \hat{H}_S + \hat{H}_T
$.
The subscripts $N$ and $S$ refer to the normal and the
superconducting electrode respectively;
the transfer of electrons through the tunnel interface
is described by the tunnel Hamiltonian
$\hat{H}_T$. The latter is expressed
in terms of quasiparticle operators $\hat{\gamma},
\hat{\gamma}^{\dagger}$ for the superconductor, and electron operators
$\hat{a},\hat{a}^{\dagger}$ for the normal metal:
\FL
\begin{eqnarray}
	\hat{H}_T
	=
	\sum _{\bf{k},\bf{p},\sigma}
	\{&&
	 t_{\bf{kp}}
	 \hat{a}^{\dagger}_{\bf{k},\sigma}
	 (u_{\bf{p},\sigma} \hat{\gamma}_{\bf{p},\sigma} +
v_{\bf{p},\sigma}\hat{\gamma}^{\dagger}_{-\bf{p},-\sigma})
\nonumber\\
	 &&+
         t_{\bf{kp}}^{\ast}
	 (u_{\bf{p},\sigma} \hat{\gamma}^{\dagger}_{\bf{p},\sigma}
		+ v_{\bf{p},\sigma}\hat{\gamma}_{-\bf{p},-\sigma})
         \hat{a}_{\bf{k},\sigma}
	\}.
\label{HT}
\end{eqnarray}
Here, $t_{\bf{kp}}$ are the tunnel matrix elements which we take to be
spin-independent, and  $u_{\bf{p},\sigma}, v_{\bf{p},\sigma}$ are the BCS
coherence factors \cite{Tinkham}; the sum is taken over momenta
$\bf{k},\bf{p}$  and spin $\sigma = \uparrow, \downarrow $.

Using second order perturbation theory in $\hat{H}_T$
one can calculate the amplitude for the transfer of two electrons
from the normal to the superconducting electrode:
\begin{equation}
	A_{\bf{k} \uparrow \bf{k}' \downarrow}
	=
	\sum _{\bf{p}} t_{\bf{kp}}^{\ast} t_{\bf{k'-p}}^{\ast} u_{\bf{p}}
v_{\bf{p}}
	\left\{
      \frac{1}{\xi_k - \varepsilon _p } +
      \frac{1}{\xi_{k'} - \varepsilon _p }
	\right\}.
\label{ampl}
\end{equation}
\narrowtext
Here the spin dependence of the coherence factors was dropped after using
the relation $v_{\bf{p}, \uparrow}=-v_{-\bf{p}, \downarrow}$.
We define electron energies $\xi_k$ and $\zeta_p$ for the normal
and the superconducting electrode respectively, and
quasiparticle energies $\varepsilon_p = \sqrt{\Delta^2 +
\zeta_p^2}$.
The denominators in (\ref{ampl}) reflect the fact that a virtual state
is formed when the first electron enters
the superconductor as a quasiparticle.
The second electron couples to this quasiparticle, thus forming a Cooper pair.
The corresponding rate $\Gamma (V)$ as a function of the voltage $V$ applied
across the junction
can be found by using Fermi's Golden Rule
\FL
\begin{equation}
	\Gamma (V)
	=
	\frac{4\pi}{\hbar} \sum _{\bf{k},\bf{k}'}
	|A_{\bf{k} \uparrow \bf{k}' \downarrow }|^2
	f(\xi _k)f(\xi _{k'})
	\delta (\xi _k + \xi _{k'} + 2eV ).
\label{andrrate}
\end{equation}
It contains the Fermi functions $f$ for
electrons with energies $\xi _k$, $\xi _{k'}$ in the normal metal.
We recall that the normal conductance of the junction is determined by the
rate $\gamma (V)$ for usual electron tunneling, which  is
proportional to $|t_{\bf{kp}}|^2$:
$
	\gamma (V)
	=
	|t_{\bf{kp}}|^2
	f(\xi _k)(1-f_r(\zeta _{p}))
	\delta (\xi _k - \zeta _p + eV )
$.

The calculation of $|A_{\bf{k} \uparrow \bf{k}' \downarrow }|^2$ in Eq.
(\ref{andrrate})
involves summations over momentum of a product of four tunnel matrixelements.
It therefore
requires an assumption about the dependence of the $t_{\bf{kp}}$
on the wave vectors $\bf{k}$ and $\bf{p}$.
This dependence is strongly related to the nature of the electron motion
in the electrodes, as we discussed above.
Following~\cite{Wilkins} we assume first that plane electron waves propagating
in the electrodes
are transmitted specularly by a rectangular tunnel barrier with, say, a length
$L_b$
and a height $U$ (See Fig. 1a). The area of the junction will be denoted by
$S$.
Specular scattering implies that the components of momentum
$\bf{k}_{\parallel}$ and $\bf{p}_{\parallel}$ parallel to the
barrier plane are conserved.
If $S$ is of the order of $\lambda _F^2$
(with $\lambda _F$ the Fermi wavelength of the electrons)
the values of $\bf{k}_{\parallel}$ and $\bf{p}_{\parallel}$ are quantized,
leading to
discrete transport channels~\cite{Imry}.
The corresponding quantum numbers are equal:
 $n_{\bf{k}} = n_{\bf{p}}$; the effective number of channels $N_{eff}$
contributing
to the transport will be calculated below.
The magnitude of $t_{\bf{kp}}$ decreases exponentially with decreasing squared
component
$\bf{k}_{\perp}^2$ perpendicular to the barrier.
This results in the assumption
$
	t_{\bf{kp}}
	\propto
	\delta _{n_{\bf{k}} , n_{\bf{p}}}
	\exp {-\bf{k}_{\parallel}^2 \lambda ^2}
$,
with
$
	\lambda
	=
	\hbar L_b/\sqrt{8mU}
$,
where $m$ is the electron mass.
The calculation of the rate~(\ref{andrrate}) is easily performed using this
model
for $t_{\bf{kp}}$, by averaging products of these matrix elements
over directions of momentum.
As a result we find that $\Gamma \propto 1/N_{eff} ^3$. Similarly we obtain
$\gamma \propto 1/N_{eff}$.
Comparing $\Gamma$ with $\gamma ^2$ we find
the effective number of transport channels penetrating the barrier, $N_{eff}$:
\[
	\frac{\Gamma}{\gamma ^2}
  	\propto
	\langle|\langle t_{\bf kp}t_{\bf k'p}
                   \rangle_{\bf p}^{}|^2\rangle_{\bf kk'}^{} /
        \left.\langle|t_{\bf kp}|^2\rangle_{\bf kp}^2\right.
	=4\pi \lambda ^2/S = 1/N_{eff}
{}.
\]

This result is obtained by assuming ballistic motion of the electrons in
the electrodes.
This assumption is correct only if the scattering of the
electron is negligible.
Scattering may occur at the boundaries of the electrodes or
at impurities inside the electrodes.
Both processes can be characterized by a space scale $l_e$, which corresponds
to
the distance the electron traverses before undergoing the first
scattering event.
Interference occurs on a space scale $\xi _{cor}$.
The ballistic picture is valid if the
typical size $\sqrt{S}$ of the junction or $\xi_{cor}$ is
smaller than $l_e$.
When these lengths are of the same order, we expect
a cross-over to different behaviour.
In the opposite limit the electron moves diffusively in the junction region.
Due to interference between incoming and backscattered electron waves $N_{eff}$
will decrease, thereby increasing the rate $\Gamma$, and hence the conductance
due to two-electron tunneling.

Now we will present a method to describe two-electron tunneling in the
diffusive
transport regime, employing the quasiclassical approximation.
This enables us to evaluate the tunnel matrix elements and express the
rate $\Gamma$ in terms of quasiclassical diffusion
propagators.
The method is similar to the one presented in Ref.~\cite{Averin}.
We start by rewriting the matrix elements
$
	t_{\bf kp}
	=
	\int dr dr' \psi _k^*(r) \psi _p(r') t(r,r')
$, where $\psi _p(r)$ forms a complete set of one-electron wave functions
in the electrodes, and
$t(r,r')$ describes the tunneling from a point $r'$
in the superconductor to a point $r$ in the normal metal (primed space
arguments refer to the superconductor).
We also define a propagator from $r_2$ to $r_1$ by
$
	K_{\xi}(r_1,r_2)
	=
	\sum _k \delta (\xi - \xi _k) \psi _k(r_1) \psi _k^*(r_2)
$.
With these definitions it is possible to rewrite Eq.~(\ref{andrrate}) as
\begin{equation}
	\Gamma (V)
	=
	\frac{4\pi}{\hbar}
	\int d\xi d\xi ' d\zeta d\zeta '
	F(\zeta;\xi, \xi ') F(\zeta ';\xi, \xi ')
	\Xi (\zeta, \zeta ';\xi, \xi ')
	f(\xi)f(\xi ')\delta(\xi + \xi ' + 2eV)
\label{andruse}
\end{equation}
with
$
	F(\zeta ;\xi, \xi ')
	=
	u(\zeta) v(\zeta)
	\left\{
		(\xi - \varepsilon)^{-1} + (\xi ' - \varepsilon)^{-1}
	\right\}
$
where $\varepsilon = \sqrt{\Delta ^2 + \zeta ^2}$, and
$$
	\Xi (\zeta, \zeta ';\xi, \xi ')
	=
	\int d^3r_1 ... d^3r_4 \int d^3r'_1 ... d^3r'_4
	t^*(r_1,r'_1) t^*(r_2,r'_2) t(r_3,r_3') t(r_4,r_4') \times
$$
\begin{equation}
	K_{\xi}(r_1,r_3)K_{\xi'}(r_2,r_4)K_{\zeta}(r_2',r_1')K_{\zeta'}(r_3',r_4')
\label{Xi}
\end{equation}
The physical meaning of Eq.~(\ref{andruse}) can be understood easily by
depicting the integrand of
Eq.~(\ref{Xi}) diagrammatically, as has been done in Fig. 2.
We see two electrons that propagate in the normal electrode with energy $\xi$
and $\xi '$.
The first electron reaches the barrier at $r_1$ and tunnels to $r'_1$,
the second electron tunnels from $r_2$ to $r'_2$; both change their energy to
$\zeta$.
In the superconductor they form a Cooper pair.
Since tunneling occurs only between neighboring positions, we have in addition
$r_i \approx r'_i$.
The diagram expresses a probability, and therefore is completed by adding the
time-reversed process.

To analyze expression~(\ref{Xi}), it is important to consider the scale of
separation of
the coordinates $r_1 \approx r'_1, ..., r_4 \approx r'_4$ lying on the
interface.
In the ballistic transport regime, these coordinates are separated only by a
few Fermi wavelengths.
In this case the contribution depends on properties of the tunnel barrier only.
Below we will concentrate on  contributions to~(\ref{Xi}) which arise when
the region of integration is defined by coordinates which are pairwise close,
but with the pairs
separated by a distance much larger than the Fermi wavelength.
These contributions contain averaged products of two propagators $K$,
which are known to decay on a mesoscopic scale
in the diffusive transport regime~\cite{Aronov}.
These products correspond to the semiclassical motion of electrons from
one point on the interface back to another point on this interface.
They describe the interference between scattered waves.
In the diffusive regime these contributions dominate; that is why we
concentrate on them.

There are three types of contributions, which are depicted in Fig 3.
Fig. 3a corresponds to the case $r_1 \approx r_2$ and $r_3 \approx r_4$.
The interference contribution originates from the normal electrode.
Fig. 3b describes the opposite situation with interference occurring in the
superconducting electrode. Here, $r_1 \approx r_3$ and $r_2 \approx r_4$.
Finally, in Fig. 3c, interference occurs both in the normal and in the
superconducting
electrode, when $r_1 \approx r_4$ and $r_2 \approx r_3$.
However, we estimated this contribution to be less important, compared to the
previous
diagrams.
Therefore, the total effect can be represented as the sum of the interference
contributions from the superconducting and the normal metal.

As an example we will discuss the contribution of Fig. 3b to the
rate~(\ref{andruse}) in some detail.
The averaged product of the propagators in the superconductor determines the
semiclassical conditional probability
$P(r'_1,r_2';n,n';t)$ that an electron with position $r_2'$ and momentum
direction $n'$ at time $t=0$
has position $r'_1$ and momentum direction $n$ at time $t$.
Since the tunnel amplitude $t(r,r')$ is nonzero only when
$r$ and $r'$ are close to the junction interface, we can restrict spatial
integrations
to planar integrations over the junction surface.
It is possible to show that
$$
	\Xi _S
	=
	\Xi _S (\zeta - \zeta ')
	=
	\frac{\hbar}{8\pi ^3 e^4 \nu _S}
	\int d^2r'_1 d^2r'_2  \int d^2n d^2n'
	g(n,r'_1) g(n',r'_2) \times
$$
\begin{equation}
	\int dt
	e^{i(\zeta - \zeta')t/\hbar}
	P(r'_1,r_2';n,n';t)
\label{Xiuse}
\end{equation}
where $\nu _S$ is the density of states for the superconductor for two spin
directions and
$\int d^2n$ denotes
integration over directions of momentum.
The function $g(n,r)$ defines the normal conductance
of the junction:
$
	G_{T}
	=
	\int d^2r
	\int d^2n g(n,r)
$.
An expression similar to~(\ref{Xiuse}) can be obtained for Fig. 3a, by
replacing
subscript $S \to N$, energies $\zeta \to \xi$, and primed space arguments by
unprimed ones.


Let us start our consideration of concrete layouts  with the simplest  geometry
of an
infinite uniform junction  between a normal and a superconducting film (Fig.
4a).
We assume that the film thickness is much less than the coherence length in the
superconductor.
Then we can exploit the picture of two-dimensional electron diffusion. The
probability
function we need is given by

\begin{equation}
P(r_1, r_2; t) = \frac{1}{4 \pi D t d} \exp {(-|r_1-r_2|^2/4Dt)},
\end{equation}
$d$ being the thickness of either the superconducting or the normal metal film.
Taking the
Fourier transform of this function with respect to time and
integrating~(\ref{Xiuse})
 over coordinates we
obtain
\begin{equation}
\Xi_S(\zeta - \zeta')=  (4 \pi^2 e^4 S \nu_S d_S)^{-1}\delta(\zeta - \zeta');
\Xi_N(\xi - \xi')=  (4 \pi^2 e^4 S \nu_N d_N)^{-1}\delta(\xi - \xi');
\label{Xifilm}
\end{equation}
The current is given by a sum of two terms ($eV \simeq \Delta \gg T$):

\begin{eqnarray}
I(V)= I_N + I_S;\nonumber \\
I_N= \frac{2 G_T^2\hbar}{e^3 S \nu_N d_N}; \nonumber \\
I_S= \frac{G_T^2 \hbar}{e^3 S \nu_S d_S} \frac{eV}{\pi \Delta
\sqrt{1-eV/\Delta}}
\label{Ifilm}
\end{eqnarray}

It is plotted in Fig. 5. The part emerging from the interference in the
normal metal does not depend on voltage. So the current sharply
jumps at zero voltage, provided $T=0$. The jump is smoothened at voltages
of the order of the temperature:
\begin{equation}
I(V,T)= I_N \tanh (eV/2T)
\end{equation}
The other contribution diverges near the threshold voltage indicating the
necessity to make use of higher order terms in tunneling amplitudes
to describe the crossover between two-electron and one-electron tunneling.

It is worthwile to compare the magnitude of the result with the one we derived
assuming ballistic motion. The order of the ratio at voltages of the order
of $\Delta$ is $ I_{int}/I_{ball} \simeq \xi_{clean} /d$ , $\xi_{clean}$
being the coherence length in the pure superconductor. Therefore the
interference
term dominates under the assumptions we made.

The coherence along a normal or superconducting  film is characterized by the
coherence length
$\xi _{cor}^{N,S} = \sqrt{D/eV}, \sqrt{D/\Delta}$, respectively.
The relations~(\ref{Ifilm}) are valid, provided the junction size is
much larger than these lengths.
In the opposite limit of small junctions, the subgap conductivity will be
determined by the
junction surroundings, rather than by the junction itself.
Let us illustrate this by
considering the geometry in Fig. 4b, where a normal electrode is connected
to a superconducting sheet by the tunnel junction.
In this case we find
\begin{equation}
\Xi_S(\zeta - \zeta')=  \frac{\hbar}{e^2} \frac{R_{\Box }^S G_T^2}{8\pi ^4 }
\ln \frac{\hbar}{(\zeta - \zeta')\tau}
\mbox{ ; }
\Xi_N(\xi - \xi')=  \frac{\hbar}{e^2}\frac{R_{\Box }^NG_T^2}{8\pi ^4 } \ln
\frac{\hbar}{(\xi - \xi')\tau}
\label{Xisheet}
\end{equation}
The time $\tau $ is of the order of $S/D$, the time spent in the junction area,
and provides a lower cut-off for the
fourier integral.
The sheet resistance of the normal (superconducting) film is given by
$R_{\Box}^{N(S)} = (e^2 \nu D_{N(S)} d_{N(S)})^{-1}$.
Indeed does the result
not only depend on the properties of the
tunnel barrier itself (through the dependence on $G_T$), but also on properties
of its surroundings
through the dependence on $R_{\Box}$.
Moreover, the dependence on the precise geometry of the layout enters through
numerical prefactors.
If, {\sl e.g.}, the tunneling would occur towards an infinite superconducting
sheet instead of
a superconducting half plane, the semiclassical probability $P$ would be twice
smaller, thus decreasing
$\Xi$, and hence the rate $\Gamma$, by a factor of $2$.
The current is again given by a sum of two terms $I_N$ and $I_S$:
\begin{equation}
	I_N
	=
	\frac{2V}{\pi} R_{\Box ,N}G_T^2 \ln \frac{\hbar}{eV \tau }
	\mbox{ ; }
	I_S
	=
	\frac{2V}{\pi} R_{\Box ,S}G_T^2 \ln \frac{\hbar}{\Delta \tau }
\end{equation}
Note that, in contrast to Eq.~(\ref{Ifilm}), the subgap conducticity depends
only
weakly on the junction area through the cut-off time $\tau$.

We finally consider the  geometry depicted in Fig. 4c.
It consists of a small island (length $L_S$, thickness $d_S$),  coupled to two
macroscopic leads by
tunnel barriers. The grain is linked  capacitively to the leads.
Electron transport through such a system, characterized by a  small
electric capacitance $C$, has been studied both experimentally and
theoretically in great detail during the past
years~\cite{Averin91Grabert}.
The key point is that variations of the charge of the island in the course of
electron  tunneling increase the electrostatic energy, typically by an amount
$E_C = e^2/2C$.  This is why
electron tunneling through a small grain is suppressed (Coulomb blockade).
The case of a superconducting island
connected to two normal electrodes (NSN geometry),
was studied recently in Refs.~\cite{Hekking,TuominenPRL1992,Eiles}.
Our method to include interference effects can also be applied to
this case.
The charging energy $E_C$
will enter our results explicitly via the virtual state
denominators in~(\ref{ampl})~\cite{Hekking}, resulting in
a dependence of the function $F$ on $E_C$.
We will restrict ourselves to the situation in which $E_C$ is smaller
than the superconducting gap: $E_C < \Delta$.
In order to calculate the contribution due to interference on the island we
assume that the time $\hbar/(\Delta - E_C)$ spent by the virtual electron on
the island is much
longer than than the classical diffusion time $L_S^2/D$.
If $\Delta \alt E_C$, the size of the island is smaller than $\xi _{cor}^S$.
In this case, the electron motion covers the whole island and the probability
$P$
is constant: $P = 1/(L_S^2d_S)$.
As a result we find:
\begin{equation}
	\Xi_S(\zeta - \zeta')
	=
	\frac{\hbar ^2 w_S G_T^2}{4\pi ^2 e^4} \delta (\zeta -\zeta')
\end{equation}
where $w_S = 1/(\nu _SL_S^2d_S) $ denotes the level spacing of the island,
which
shows once more that the rate~(\ref{andruse}) is not only determined by
properties
of the tunnel barrier.
The corresponding current $I_S$ reads
$$
	I_S
	=
	V \frac{4 \hbar}{\pi e^2}G_T^2 \frac{w_S \Delta }{E_C^2} \times
$$
\begin{equation}
	\left[
	\frac{\pi}{2} - \frac{2\Delta }{\sqrt {\Delta ^2 - E_C^2}}
		\left\{
	1 - \frac{E_C^2}{\Delta ^2 - E_C^2}
	\right\}
	\arctan \sqrt { \frac{\Delta - E_C}{\Delta + E_C}}
	+
	\frac{\Delta E_C}{\Delta ^2 - E_C^2}
	\right]
\end{equation}
When interference in the normal electrode is taken into account, we find
$
	\Xi_N(\xi - \xi')
	=
	(\hbar/e^2)(R_{\Box }^N G_T^2/4\pi ^4) \ln [\hbar/(\xi - \xi')\tau ]
$, like in Eq.~(\ref{Xisheet}), however larger by a factor of 2, due the
difference in geometry.
The resulting current reads:
\begin{equation}
	I_N
	=
	\frac{4V}{\pi ^3} R_{\Box }^N G_T^2
	\ln \frac{\hbar}{eV\tau }
	\left\{
	\frac{4\Delta}{\sqrt{\Delta ^2 - E_C^2}}
	\arctan {\sqrt{\frac{\Delta + E_C}{\Delta - E_C}}}
	\right\}^2
\end{equation}

In conclusion, we evaluated the low-voltage supgap conductivity of NS boundary
interface.
In many interesting cases it was shown to be determined by the conditions of
electron
motion in the electrodes rather than the properties of the tunnel barrier.
Therefore
it depends on the system layout on the mesoscopic scale. We presented an
approach
which gives exact results for any layout given.

The authors are grateful to  M. Devoret, D. Esteve, J. Mooij, A.Schmid, and G.
Sch\"on for
a set of valuable discussions.

\figure{Two typical realizations of a tunnel junction between a normal (N)
and a superconducting (S) electrode.
In (a) the electron moves ballistically, in (b) diffusively in the junction
region.}
\figure{Diagram corresponding to Eq.~(\ref{Xi}).
Electrons propagate (solid lines) with energy
$\xi, \xi'$
in the normal (N) and energy $\zeta, \zeta'$ in the superconducting (S)
electrode.
They tunnel through the barrier (shaded region) at postitions
$1,...,4$, marked by crosses.}
\figure{Contributions to the subgap conductivity due to interference in
(a) the normal electrode, (b) the superconducting electrode, and (c) both
electrodes.}
\figure{Various layouts discussed in the text: (a) infinite uniform junction,
(b)
normal electrode connected to a superconducting halfplane, and (c) a
superconducting
island connected to two normal electrodes.}
\figure{$I-V$-characteristcs for an infinite junction. The curves (from bottom
to top)
represent $I_S$, $I_N$, and $I(V)=I_S +I_N$}

\end{document}